\documentclass[preprint]{iucr}              

     \journalcode{J}              
\usepackage{graphicx}
\usepackage[T1]{fontenc}
\usepackage[utf8]{inputenc}
\usepackage{amsmath}

\begin{document}                  



\title{Application of signal separation to diffraction image compression and serial crystallography}


\cauthor[a]{Jérôme}{Kieffer}{jerome.kieffer@esrf.fr}{}
\author[a]{Julien}{Orlans}
\author[a]{Nicolas}{Coquelle}
\author[a]{Samuel}{Debionne}
\author[b]{Shibom}{Basu}
\author[a]{Alejandro}{Homs}
\author[a]{Gianluca}{Santoni}
\author[a]{Daniele}{De Sanctis}

\aff[a]{European Synchrotron Radiation Facility;71, avenue des Martyrs;CS 40220;38043 Grenoble Cedex 9 \country{France}}
\aff[b]{EMBL Grenoble; 71 avenue des Martyrs; CS 90181; 38042 Grenoble Cedex 9; \country{France}}









\maketitle                        

\begin{synopsis}
Precise background assessment and application to single crystal image compression and serial crystallography data pre-processing. 
\end{synopsis}

\begin{abstract}

We present here a real-time analysis of diffraction images acquired at high frame-rate (925 Hz) and its application to macromolecular serial crystallography.
The software uses a new signal separation algorithm, able to distinguish the amorphous (or powder diffraction) component from the diffraction signal originating from single crystals. 
It relies on the ability to work efficiently in azimuthal space and derives from the work performed on pyFAI, the fast azimuthal integration library.
Two applications are built upon this separation algorithm: a lossy compression algorithm and a peak-picking algorithm; the performances of both is assessed by comparing data quality after reduction with XDS and CrystFEL.

\end{abstract}


\section{Introduction}

X-ray macromolecular crystallography is one of the most successful methods to determine the atomic structure of biological molecules. 
The achievable diffraction quality may often be limited by radiation damage. 
Although cryogenic conditions permit to extend the lifetime of crystals in the X-ray beam and to increase the maximum absorbed dose before inducing damage, they may hinder physiologically relevant conformations. 
This limitation has renovated the interest for room temperature macromolecular crystallography by applying a more drastic approach to overcome the radiation damage problem by collecting data from thousands of small crystals, in what became known as serial crystallography. 
First developed at X-ray free electron laser sources (\textsc{xfel}) \cite{Chapman2011, structure_sx} the method is currently applied also at synchrotron sources \cite{ssx, ssx_desy, ssx_id13, ssx_diamond}.


\subsection{Serial crystallography using synchrotron sources (SSX)}
%
Serial crystallography consists in exposing thousands of small crystals to the X-ray beam only once in a serial way.
Diffraction is collected with a very high flux density, in order to extract the most information from a single shot.
This is in contrast with traditional rotational crystallography, where a complete dataset is collected from a single crystal rotated around one (or several) axis. 
Those SSX images represent a slice through the reciprocal space, thus intersect a lower fraction of the reciprocal space as the crystal is still, in comparison with rotational crystallography.
To achieve a complete dataset, thousands of frames have to be collected, individually indexed and then merged.
The high flux needed to collect all diffraction signal from a single crystal within a single exposure makes the SSX technique a good candidate to benefit from the 4\textsuperscript{th} generation synchrotron sources, such as the new ESRF-EBS update \cite{EBS}.
However, macromolecular crystallography beam-lines (MX) have been extremely specialized towards rotational data-collection and thus require modifications to the experimental setup to perfom SSX experiments.
The synchrotron serial crystallography beamline (ID29 at the ESRF) has been built to have a dedicated environment to perform SSX experiments with a high flux (using a larger energy bandwidth with a multilayer monochromator), a high-speed chopper (to separate X-ray pulses), several sample delivery methods and a fast detector.

\subsection{\textsc{Jungfrau} 4M detector}
Macro-molecular crystallography has enormously progressed in the last decades with to the introduction of photon-counting detectors \cite{pilatus}. 
With their absence of read-out noise and their fast speed, the main limitation of photon-counting detectors is the achievable count-rate, i.e. how fast the electronic of a pixel is able to count arriving photons.
Unlike photon-counting detectors such as the Eiger detector \cite{Eiger}, the \textsc{Jungfrau} detector \cite{jungfrau2016} is an integrating detector, thus, it is not limited by the count rate, even under the very intense flux expected when recording Bragg-peaks.
To cope with this photons density, every single pixel implements an automatic gain switching mechanism (3 levels) which offers a precision of the order of one third of a keV on the higher gain mode, a precision of the order of one photon in the intermediate level and the ability to cope with thousands of photons in the lower gain mode.
Moreover, the \textsc{Jungfrau} is able to operate at two kilohertz, which means the image must be read every half millisecond.
Since the \textsc{Jungfrau} detector is an integrating detector, dark current and flat-field correction have to be applied: every pixel has three, so called, pedestals and three gain values (one for each gain level). 
This large number of parameters per pixel makes the pre-processing of raw signal challenging at full speed: the signal from a single pixel, initially packed with 16 bits per pixel, gets expanded to 32 bits (floating point or integer value), doubling the required bandwidth and the storage size \cite{jungfrau_PSI}.
The ID29 beamline features a \textsc{Jungfrau} 4M detector, operating at 1kHz, pace imposed by the chopper and synchronized with the photon bunches from the ESRF.
At nominal speed, the detector will produce 16 GBytes of pre-processed data per second, making the data analysis and storage extremely challenging.

\subsection{Requirements for online data processing}

Serial crystallography is one of the techniques where online data processing is likely to have most impact:
millions of images are to be collected, and existing detector already saturate the fastest storage systems, not even considering the cost of this storage.
Beside this, only a few percentage of the frames are expected to contain diffraction signal and out of them, a fraction will be indexed, integrated and thus useful to solve the protein structure.
Efficient processing of raw images is thus essential for SSX, with 4 levels of increasing complexity:
\begin{itemize}
    \item image reconstruction with pedestal correction
    \item \textit{veto-}algorithm: be able to sieve-out images with poor signal
    \item save only pixels with diffraction signal
    \item precise location of peak position with indexation \cite{toro}
    \item real-time integration of diffraction peaks
\end{itemize}.

Reconstruction and pedestal correction have been already described in \citeasnoun{lima2}.
This document will focus on the subsequent steps: the detection of signal is addressed in section 2, sparse data compression in section 3, and peak-finding in section 4 before drawing some conclusions in section 5.


\section{Algorithm for the separation of the amorphous background from the Bragg peaks}
\subsection{Background scattering}
The simplest implementation of Bragg peaks separation is to assume that the background signal originates from scattering of amorphous material 
giving an isotropic signal that ideally presents only smooth variations.
Before background subtraction, the raw signal has to be corrected for dark noise and for any systematic anisotropic effects such as polarization corrections.
Unlike X-FEL, synchrotron X-ray beam is better characterized in energy and shows little to none pulse to pulse variability.
All anisotropic correction can be easily modelled and taken into account.
The same method can be extended to separate Bragg peaks from powder diffraction if the powder signal is isotopic, i.e. without preferred orientation. 

The initial implementation of signal separation in pyFAI \cite{pdj2013} was relying on a 2D polar transform followed by a median filter in the azimuthal dimension to calculate the amorphous scattering curve.
Although this method has been successfully used for large dataset analysis \cite{brocades}, it presents four major drawbacks:
\begin{itemize}
\item The 2D averaging mixes the signal originating from several pixels and blurs the signal. 
\item Pixel-splitting is needed to leverage the Moiré effect in the 2D averaging, but this further increases the blurring \cite{moire}. 
\item The 1D curve obtained after the application of the median filter shows sharp jumps from one azimuthal bin to its neighbour.
\item The median filter is computationally heavy since it is required to sort out every azimuthal bin.
\end{itemize}

We improved on this by developing a new, efficient way of performing the azimuthal averaging (including the associated uncertainty propagation). 

\subsection{Efficient azimuthal averaging and uncertainties evaluation}

\subsubsection{Pre-processing:}
The first step of the analysis consists in applying a pixel-wise correction for dark current and several normalization corrections \cite{pyfai_2020}:
\begin{equation}
\label{norm}
I_{cor} = \frac{signal}{norm}  = \frac{I_{raw} - I_{dark}}{F \cdot
\Omega \cdot P \cdot A \cdot I_0} 
\end{equation}
In  equation \ref{norm}, the numerator (referred as \textit{signal} hereafter) is the subtraction of the dark current $I_{dark}$ from the the detector's raw signal $I_{raw}$.
The denominator (hereafter \textit{norm}) is a normalization factor composed of the product of  $F$:  a factor accounting for the flat-field correction, $\Omega$: the solid angle subtended by a given pixel, $P$: the polarisation correction term and $A$: the detector's apparent efficiency due to the incidence angle of the photon on the detector plane. 
For integrating detectors, photons with high energy see longer sensor path with larger incidence angles compared to the normal thickness, thus they have a higher detection probability.
Intensity is also normalized by the incoming flux $I_0$, but being it independent from the pixel position, this correction can be applied when convenient.

\subsubsection{Azimuthal averaging:} 

Historically, the azimuthal averaging was implemented using histograms \cite{fit2d1996}.
Since the geometry of the experimental setup is fixed during the acquisition, a look-up table listing all pixels contributing to each  azimuthal-bin can be built and used to speed-up calculations \cite{pyFAI_gpu}.
The azimuthal transformation being a linear transformation, it can be implemented as a matrix multiplication, with a sparse-matrix representing the transformation and a dense vector containing the flattened view of the diffraction image. 
The compressed sparse row (CSR) matrix representation is preferred for its efficiency in performing dot products with dense vectors \cite{SpMV}.
The coefficients $c_{i,r}$ of the matrix are the fraction of area of a pixel $i$ falling into the radial bin $r$.
In the case where pixel splitting is deactivated these coefficients  ($c_{i,r}$) are always one (and zero elsewhere) since each pixel contributes to a single bin.
The sparse matrix multiplication can be used to sum efficiently values for all pixels belonging to the same bin.
The summed signal divided by the summed normalization provides the weight-averaged intensity over all pixels falling in the bin at the distance $r$ from the center, as formalized in equation \ref{avg}: 
\begin{equation}
\label{avg}
<I>_{r} = \frac{\sum\limits_{i \in bin_r} c_{i,r} \cdot signal_i}
                        {\sum\limits_{i \in bin_r} c_{i,r} \cdot norm_i} 
\end{equation}  

\subsubsection{Uncertainty evaluation from Poisson distribution.}
Photon counting detectors, such as Eiger detectors, suffer from hardly any error beside the counting uncertainty which is often referred to as Poisson statistics.
This statistical law is described by a single parameter $\lambda$ which is related to the mean $\mu$ and standard deviation $\sigma$ from a normal distribution by: $\lambda=\mu=\sigma^2$.
Other sources of noise, like the dark current noise in the case of an integrating detector,  superimpose quadratically on the Poisson noise for integrating detectors, as presented in equation \ref{poisson}:     
\begin{equation}
\label{poisson}
var_I = (\sigma_I)^{2} = <I_{raw}> + (\sigma_{dark})^{2}  
\end{equation}

During the azimuthal integration, the coefficient of the sparse matrix needs to be squared at the numerator when propagating the variance (equation \ref{varianceP}) to have uncertainties $\sigma$ proportional to the fraction of the pixel considered.
\begin{equation}
\label{varianceP}
(\sigma_{r}(I))^2 = \frac{\sum\limits_{i \in bin_r} c_{i,r}^2 \cdot \sigma_i^2}
                  {\sum\limits_{i \in bin_r} c_{i,r} \cdot norm_i} 
\end{equation}
One should distinguish the \textit{uncertainty of the mean} (sometimes referred to as the standard error of the mean, $sem$), 
which describes the precision with which the mean is known (and described in \citeasnoun{pyfai_2020}),
from the \textit{uncertainty of the pixel value} (often referred to as standard deviation, $std$) which describes the uncertainty with which the pixel value is known. 
Those two value differ only by the square root of the number of measurements in the case of an arithmetic mean: $sem = std/\sqrt{N}$ with $N$ being the number of pixel contributing to the bin.
When considering the weighted average, the previous formula becomes:
\begin{equation}
\label{sem}
sem_r = std_r \frac{\sqrt{\sum\limits_{i \in bin_r} c_{i,r}^2 \cdot norm_i^2}}{\sum\limits_{i \in bin_r} c_{i,r} \cdot norm_i}
\end{equation}
Thus, the more data points are collected, the more precisely the mean value is known but the uncertainty for a given point remains the same.
Since this document focuses on the uncertainties of pixel values, the \textit{standard deviation} will systematically be used from here on.  

\subsubsection{Uncertainty evaluation from the variance in a bin.}

Unlike photon counting detectors, most detectors do not follow the Poisson distribution and therefore the definition of a relation $\sigma^2 = f(I)$ is not simple, if possible at all. 
The integrating \textsc{Jungfrau} detector has a complex gain switching mechanism \cite{jungfrau_PSI} which makes this equation complicated.
A generic approach is proposed to measure the variance in every single azimuthal bin:

When considering the diffraction of an isotropic compound (liquid, amorphous or perfect powder), all pixels contributing to the same radial bin should see the same flux of photons (after correction of anisotropy like polarization) and the deviation to their intensities can be used to estimate the uncertainty. 
This approach is of course limited when considering the signal coming from few large crystalites (where rings becomes spotty) but it provides an upper bound for the uncertainty.
Variances (thus standard deviations) are usually obtained in a two steps procedure: one pass to calculate the average value (equation \ref{avg}) and a second to calculate the deviation to the average (equation \ref{varA}). 
This double pass approach can be implemented using sparse matrix multiplication. 
It requires twice the access to each pixel value, and extra storage space, but it is numerically robust (i.e. not prone to numerical-error accumulation).
\begin{equation}
\label{varA}
    (\sigma_{r}(I))^2 = \frac {\sum\limits_{i \in bin_r} c_{i,r}^2 \cdot norm_i^2 \cdot (\frac{signal_i}{norm_i}-<I>_r)^2}
                              {\sum\limits_{i \in bin_r} c_{i,r}^2 \cdot norm_i^2}
\end{equation}
and:
\begin{equation}
\label{varB}
(\sigma_{r}(<I>))^2 = \frac{\sum\limits_{i \in bin_r} c_{i,r}^2 \cdot norm_i^2 \cdot (\frac{signal_i}{norm_i}-<I>_r)^2}
                         {(\sum\limits_{i \in bin_r} c_{i,r} \cdot norm_i)^2}  
\end{equation}

Single pass implementations of variance calculation are faster than double pass ones since they access pixels only once and offer, in addition, the ability to perform parallel reductions \cite{Blelloch}, i.e. work with blocks of pixels.
\citeasnoun{variance2018} present a complete review on the topic, which introduces a formalism adapted here for crystallography.
Assume the weight for a pixel being $\omega_i = c_i \cdot norm_i$.
If $P$ is a partition of the ensemble of pixels falling into a given azimuthal bin, let $\Omega_{P}$, $V_{P}$ and $VV_{P}$  
be the sum of weights (Eq \ref{omega}), the weighted sum of $V$ (Eq. \ref{Vp}) and the weighted sum of deviation squared (Eq \ref{VVp}) over the partition $P$: 
\begin{equation}
\label{omega}
\Omega_{P} = \sum\limits_{i \in P} \omega_i = \sum\limits_{i \in P} c_i \cdot norm_i 
\end{equation}
\begin{equation}
\label{omegaomega}
\Omega\Omega_{P} = \sum\limits_{i \in P} \omega_i^2 = \sum\limits_{i \in P} c_i^2 \cdot norm_i^2 
\end{equation}
\begin{equation}
\label{Vp}
V_{P} = \sum\limits_{i \in P} \omega_i \cdot v_i =  \sum\limits_{i \in P} c_i \cdot signal_i
\end{equation}
\begin{equation}
\label{VVp}
VV_{p} = \sum\limits_{i \in P} \omega_i^2 \cdot (v_i - V_{P}/\Omega_{P})^2 
\end{equation}

The weighted average and associated variances are then expressed as:
\begin{equation}
\label{weighted_avg}
<I>_P = \frac{V_{P}}{\Omega_{P}} =  \frac{\sum\limits_{i \in P} c_i \cdot signal_i}
                        {\sum\limits_{i \in P} c_i \cdot norm_i} 
\end{equation}
\begin{equation}
\label{std2}
std^2 = (\sigma_P(I))^2 = \frac{VV_{P}}{\Omega\Omega_{P}}
\end{equation}
\begin{equation}
\label{sem2}
sem^2 = (\sigma_P(<I>))^2 = \frac{VV_{P}}{\Omega_{P}^2}
\end{equation}

In this formalism, equations \ref{avg} and \ref{weighted_avg} on one side and equations \ref{varA} and \ref{std2} are actually equivalent.
\citeasnoun{variance2018} present the way to perform the union of two sub-partitions $A$ and $B$ of a larger ensemble which opens the doors to parallel reductions, which are especially efficient when implemented on GPU:
\begin{equation}
\label{union_omega}
\Omega_{A \cup B} =  \Omega_{A} + \Omega_{B} 
\end{equation}
\begin{equation}
\label{union_valeur}
V_{A \cup B} =  V_{A} + V_{B} 
\end{equation}
\begin{equation}
\label{union_variance1}
VV_{A \cup b} =  VV_{A}  +  \omega_b^2(v_b-\frac{V_A}{\Omega_A})(v_b-\frac{V_{A\cup b}}{\Omega_{A\cup b}})
\end{equation}
\begin{equation}
\label{union_variance2}
VV_{A \cup B} \approx  VV_{A} + VV_{B} +  \frac{\Omega\Omega_{B}(V_A\cdot \Omega_B-V_B\cdot \Omega_A)^2}{(\Omega_{A\cup B}.\Omega_A.\Omega_B^2)}
\end{equation}
While equations \ref{union_omega} and \ref{union_valeur} are trivial, equation \ref{union_variance1} describes the nominator of the variance of an ensemble when adding an extra member $b$ to the $A$.
Unfortunately, the slight difference of formalism between \citeasnoun{variance2018} and this work prevent some simplification to occur and leads to the approximate numerator of the variance ($VV$) in the case of the union of two ensemble $A$ and $B$ (eq. \ref{union_variance2}), used in OpenCL reduction \footnote{see https://github.com/silx-kit/pyFAI/blob/main/doc/source/usage/tutorial/Variance/uncertainties.ipynb}.
A numerical stability issue can arise from it when $V_A$ or $V_B$ are very small and this issue is addressed by using double-precision arithmetic when implemented on CPU and double-word arithmetic when running on GPU \cite{double_word}.

\subsubsection{Comparison of uncertainty models:}
\begin{figure}
\label{fig_std}
\begin{center}
\includegraphics[width=14cm]{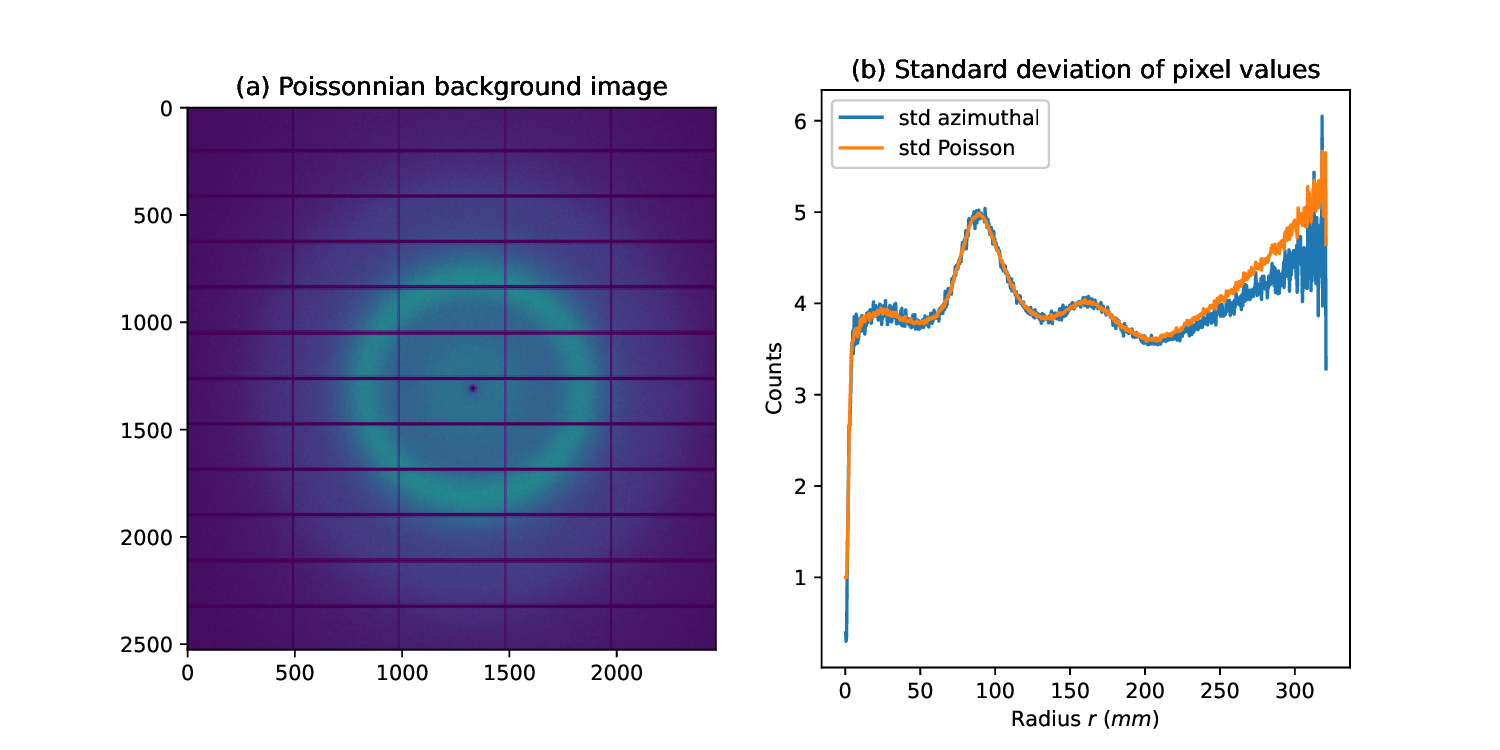}
\caption{(a) Simulated diffraction frame with pure azimuthal Poisson noise of a Pilatus 6M detector and (b) uncertainties for pixel intensity as measured with the distance to the mean (azimuthal-model, blue) or from the Poisson model (orange).}
\end{center}
\end{figure}

Figure \ref{fig_std} (right hand side) presents the uncertainties (for the pixel value) as calculated from a background frame with pure Poisson noise (displayed on the left hand-side, synthetic data) using the two algorithms previously described: the Poisson model or calculated from the variance in the azimuthal bin.
While the two curves show similar amplitude, except in the corner of the detector where very few pixels contribute to each of the azimuthal bin, the variability of the ``azimuthal'' model is much more important from one bin to the neighboring one.

\subsection{Histogram intensity }

\begin{figure}
\label{fig2}
\begin{center}
\includegraphics[width=14cm]{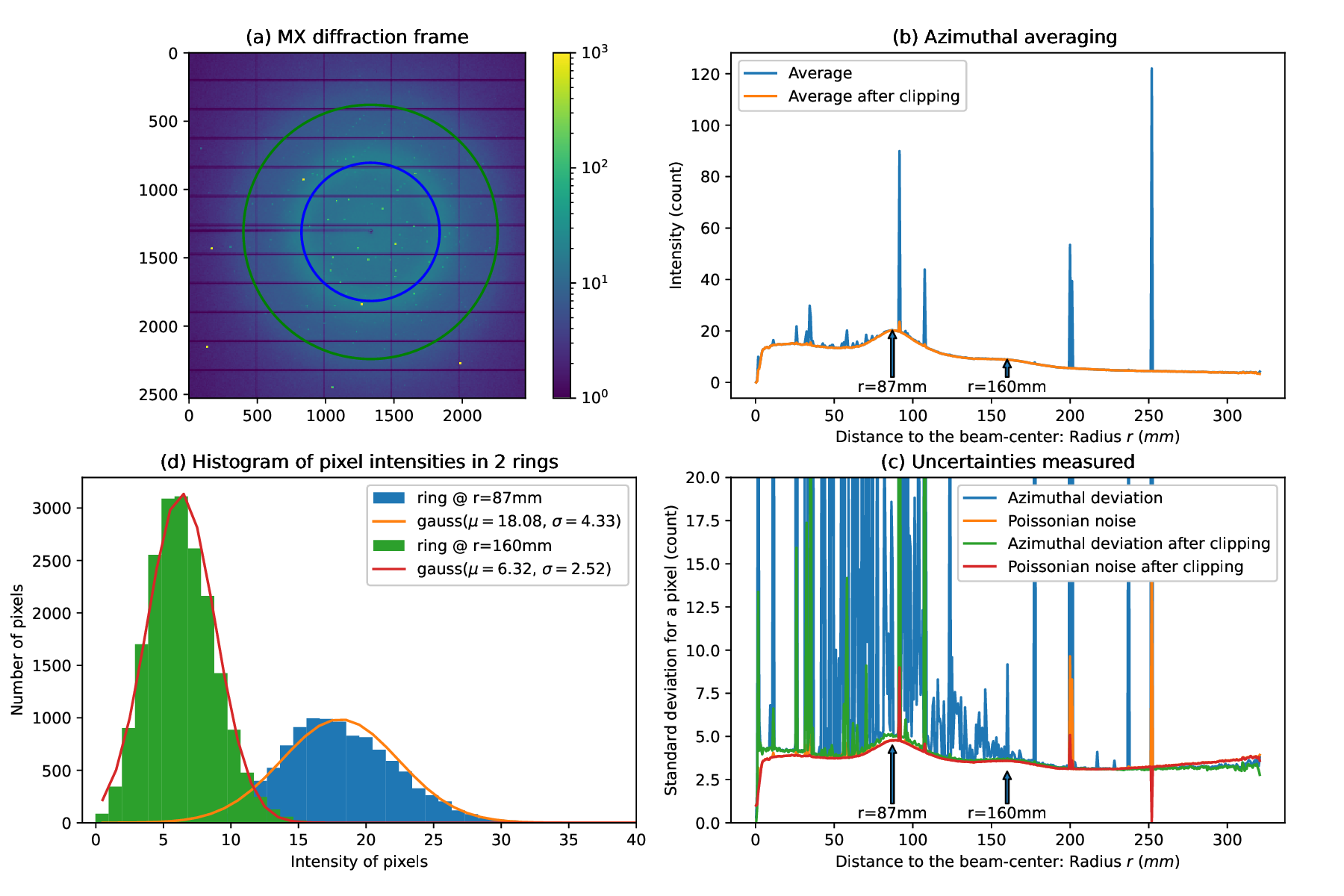}
\caption{Single crystal diffraction frame obtained from insulin with a Pilatus 6M (a) with the azimuthally averaged signal (b), 
before and after clipping data. Uncertainties are presented in (c) when calculated assuming a Poissonian error model (orange, red) or when measuring the deviation within all pixels in a ring (green, blue).
(d) Histogram of intensities for two rings at r=87mm and r=160mm from beam center with the distribution fitted as Gaussian curves: $g(x)\propto exp(-\frac{(x-\mu)^2}{2\sigma^2})$.}
\end{center}
\end{figure}

The figure \ref{fig2} presents the diffraction from a single crystal of insulin collected with a Pilatus 6M detector (Fig \ref{fig2}a) and several curves obtained from azimuthal integration of those data: 
figure \ref{fig2}b is the azimuthally integrated signal (blue curve) where Bragg peaks are seen as spikes on top of a smooth background.
Figure \ref{fig2}c presents the uncertainties measured according to the Poisson distribution (orange curve) 
or the deviation in the ring (blue curve). 
The later presents much larger values since Bragg peaks contribute a lot to the deviation despite they represent few pixels: this highlights the sensitivity of the mean/std to outliers.
In the figure \ref{fig2}d are presented histograms of pixel intensity for pixels laying at 87mm and 160mm from the beam center. 
Each of those histograms is composed of a bell-shaped distribution with 
few positive outliers which are usually Bragg peaks.   
Those histograms in figure \ref{fig2}d have been fitted with a Gaussian curve and both the center ($\mu$) and the width ($\sigma$) of the curve match 
roughly with the average (in \ref{fig2}b) and uncertainties (in \ref{fig2}c).  

The core idea of the algorithm for background extraction is to model the distribution of background pixels. Unlike Bayesian statistics \cite{Sivia2006} where the cost function is usually tuned to weight less outlier, here, those outliers are simply flagged and discarded.
Positive outliers can reasonably be assigned to Bragg peaks and negative outliers to shadows or defective pixels. 
The distribution is recalculated  after discarding pixels for which intensity differ from the average value by more than $n$ times the standard deviation (Eq. \ref{clip}), where $n$ is called the Signal to Noise Ratio (SNR).
This clipping re-centers the distribution of remaining pixels since the mean is sensitive to outlier values:
\begin{equation}
\label{clip}
|I - <I>| > n \cdot \sigma(I)
\end{equation}
The orange plot in figure \ref{fig2}b presents the average after having discarded those outliers, and the orange and green curve of figure \ref{fig2}c are the uncertainties calculated after this clipping. 
After clipping, the average and uncertainties curves have lost most of their spikes, which means that most Bragg peaks and shadowed pixel have been discarded.

Of course this works only if the background signal is isotropic, ideally smoothly varying, and there are many more background pixels than peaks or shadowed pixels. 
While shadows can be handled with a user defined mask, anisotropy in the background scattering, as it is sometimes observed with certain stretched plastic films in fixed-target mode, is much more challenging and will not be addressable with this analysis.   

\subsection{Sigma-clipping}
The \textit{sigma-clipping} algorithm consists in applying this outlier rejection several times.
If the initial distribution is mono-modal, this algorithm enforces gradually the data to be sampled symmetrically around the maximum probability, which is likely to look like a normal distribution.
If the initial distribution is more complicated (typically multi-modal), the necessary larger standard-deviation will prevent most outlier pixels from being rejected, making it more conservative.
The sigma-clipping algorithm takes two parameters: the number of iterations and the rejection cut-off (SNR).
Despite the execution time is proportional to the number of iteration of sigma-clipping, iterations should continue until no more outliers are found, so that the background data can be treated assuming a normal distribution. 
Since the loop exits as soon as no more outliers were discarded at the clipping step, having an arbitrary large number of iteration is not really an issue for the execution time and the number of actual iteration is usually few (3 is commonly observed).       

\subsubsection{Limits of the Poissonian approach:}

The evaluation of uncertainties based on the variance within a radial shell (azimuthal model) has been developed after numerical artifacts were discovered while performing sigma-clipping with a Poissonian approach. 
Some azimuthal bins were showing no pixel contribution at all and thus appeared without any mean nor uncertainties, jeopardizing the complete background extraction algorithm. 
This artifact was directly linked to the usage of Poisson statistics and can be demonstrated with a simple distribution of 2 pixels, having values 1 and 199. 
The mean of this distribution is 100 and the standard deviation is also close to 100 while the uncertainty derived from a Poissonian law would  be close to 10 (i.e. $\sqrt{100}$). 
With the azimuthal error model, both pixels are a $1\sigma$ from the mean while with the Poissonian error-model, pixels are at $10 \sigma$.  
This explains why bins featuring strong Bragg-peaks on top low background got completely emptied from any contributing pixels when sigma-clipping was performed assuming Poissonian noise.

Unlike the Poisson error-model, the uncertainties obtained from the azimuthal-model are resilient to diffraction data coming from several types of samples but show much more variability from one bin to its neighbor (Fig. \ref{fig_std}b). 
PyFAI introduces an \textit{hybrid} error-model which uses the azimuthal error-model for the sigma-clipping stage which trims the ensemble of pixels to become mono-modale.
The uncertainties are then calculated using a the Poisson error-model on the trimmed ensemble.

\subsubsection{Clipping threshold}
can be automatically calculated based on a variation on Chauvenet's criterion \cite{chauvenet} where one would accept to discard only a single pixel in a ring with a signal already following a normal law. 
Thus, the threshold value is adapted to the $size$ of the distribution, i.e. the number of pixels in each ring (Eq. \ref{chauvenet}), which can reach several thousands and shrinks with iterations.
Typically the numerical value for this cut-off varies from 2 to 4.   
\begin{equation}
\label{chauvenet}
SNR_{chauv.} =  \sqrt{2 log(\frac{size}{\sqrt{2 \pi}})}
\end{equation}

The worse case scenario for sigma-clipping corresponds to an initial distribution very far from a normal distribution, as the bimodal distribution seen in the previous paragraph.
Another challenging situation occurs close to the detector corners where the background signal is low and the size of the distribution is decreasing. 
For example this cut-off parameter increases from 2.7 to 3.5 when the size of the ensemble increases from 100 to 1000 elements.
So, for a Poissonian detector and a low count rate of one ($\lambda = \mu = \sigma^2 = 1$), any pixel with intensity greater than 4 is discarded with the ensemble of 100,  greater than 5 for the ensemble of 1000 pixels.

\section{Application to lossy image compression for X-ray diffraction}
Diffraction images from protein crystals usually exhibit an isotropic background on top of which Bragg peaks appear (discarding any diffuse scattering). 
The sigma-clipping algorithm can be used to select the background level and more importantly the associated uncertainty.
This lossy compression algorithm consists in saving only pixels which intensity is above the average background value ($\mu$) plus $n$ standard deviation ($\sigma$). 

\begin{figure}
\label{distribution}
\begin{center}
\includegraphics[width=9cm]{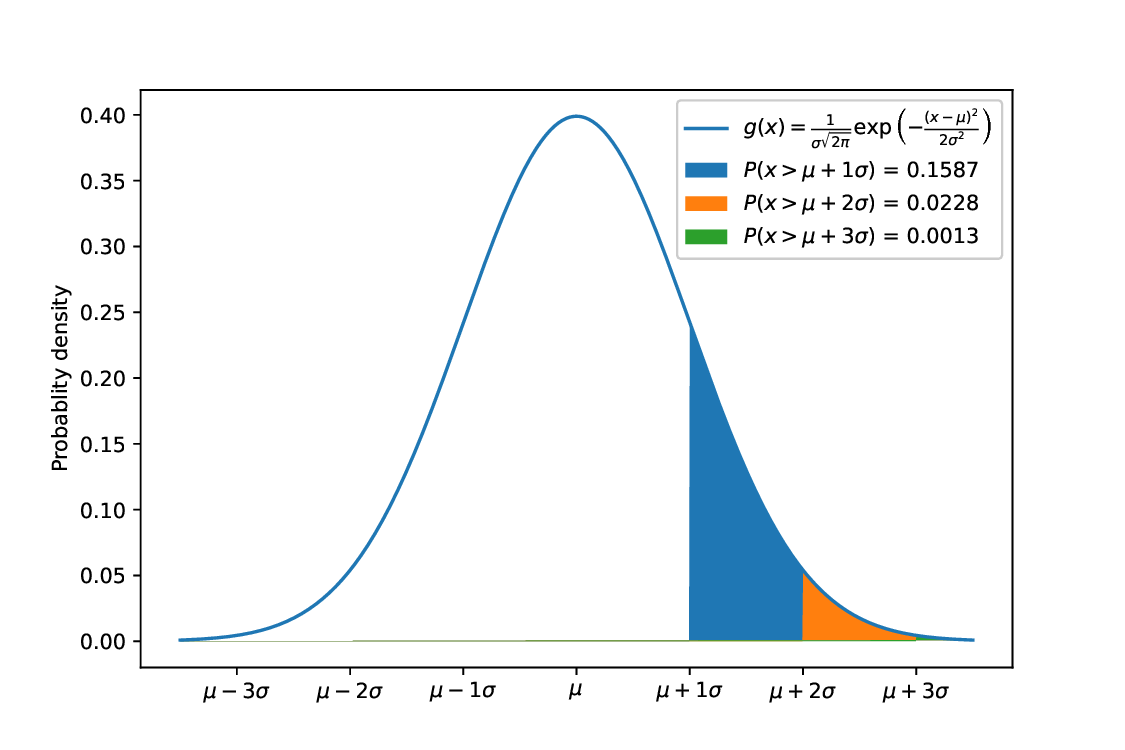}
\caption{Normal distribution and probability of having pixels with intensities above a certain threshold.
The cut-off parameter governs how much signal is integrally kept, thus the achievable compression rate on the one hand and the limit of quality of data on the other. It is the user's responsibility to set wisely this threshold.}
\end{center}
\end{figure}

The decompression simply restores those intense pixels and builds a smooth background for the missing ones (possibly with noise).
The cut-off value $n$ (also called $SNR_{pick}$) controls the amount of data to store. 
It is linked to the compression ratio: assuming a normal distribution has been enforced at the sigma-clipping stage: 16\% of the pixel are to be recorded with $n=1$;  2.3\% for $n=2$ and only 0.13\% for $n=3$ as depicted in figure \ref{distribution}.

The compressed data-format consists in:
\begin{itemize}
    \item Pixels intensities and positions for pixels worth saving.
    \item Average level of the background and associated uncertainties as function of the distance to the beam-center.
    \item The distance to the center of every pixel.
\end{itemize}
Since diffraction analysis software are performing some kind of noise-level analysis, the background signal has to be regenerated with intensity and noise similar to the original data.

\subsection{Sparsification / Compression}

The sigma-clipping algorithm was originally written for real-time sparsification of single crystal diffraction data and its integration into the LImA2 detector control system \cite{lima} for the \textsc{Jungfrau} 4M detector used at ESRF ID29 is described in \cite{sri2021}.
The real-time constrain imposed to develop code running on GPU since those devices are several times
faster than equivalently priced processors.
All algorithms were developed in OpenCL \cite{opencl_khronos} and implemented in the \textit{pyFAI} software package (MIT-licence, available on github).
A command line tool called `sparsify-Bragg' is available for testing off-line since version 2023.01.

All the pixel coordinates and intensities are stored in a HDF5 container \cite{hdf5} following the NeXus  convention \cite{nexus}, together with a snippet of Python code explaining how to rebuild the dataset.
All sparse datasets (averaged and uncertainties curves, pixel coordinates, etc..) are compressed with the bitshuffle-LZ4 \cite{bitshuffle} lossless compression.

\subsection{Densification / Decompression}
Since no crystallographic software package can deal with this sparse format (yet), the densification code was developed to regenerate initial frames and the `densify-Bragg` program was made available as part of the FabIO \cite{fabio} software package (MIT license). 
The source code is available on github and `densify-Bragg` is available since version 2022.12, available via usual channels like pip or conda install. 
The software constrains for this densification code are very different from the one for sparisfication since this code can be used by the final users.
For this reason `densify-Bragg' was optimized to run on multi-core CPU.
Maybe an important consideration is whether, regardless from the file format, it is necessary to reconstruct the background or not. 
In fact, some crystallographic reduction program like CrysAlisPro \cite{crysalis} provide a better result with noise-less background while XDS \cite{xds}, which performs a deep noise analysis, needs to have the noisy background properly restored.
Shaded regions are never reconstructed properly and should be masked adequately in the reduction software.

Future development will focus on a HDF5-plugins able to provide access to densified images from their sparse representation using user-defined function in HDF5 \cite{hdf5-udf}.
This will allow any analysis software (already able to read HDF5 files) to treat sparse data as if they were dense, removing the burden of densifing the images from the users.

\subsection{Performances on a single crystal protein dataset}
The performances for a lossy compression algorithm are to be evaluated along many directions: compression and decompression speeds, compression ratio and degradation of the recorded signal.
In the following example we present the sparsification of a Lysozyme (HEWL) dataset obtained using a traditional oscillation data-collection. 
Data were collected on an Eiger 4M detector \cite{lysozyme}, selected for its similarity in size and performances with the \textsc{Jungfrau} detector.
Those data are then densified again to regenerate the data and processed in XDS \cite{xds}.
Data quality indicators are finally compared between the original dataset and the one which went through the lossy compression presented here.


\subsubsection{Compression ratio} 
After sparsification (picking cut-off: $2\sigma$, error-model: Poisonnian), the dataset still weights 103MB which represents a 15x compression ratio compared to the standard procedure.
For conformance with the state of the art, the reference dataset was re-compressed using the bitshuffle-LZ4 algorithm \cite{bitshuffle}, for which the 1800 frames weight 1500MB (instead of the 5000GB of the original files compressed in LZ4).

The maximum theoretical compression ratio for $2\sigma$ is 22x (figure \ref{distribution}, neglecting the storage of the background data and effects of the lossless compression).
To evaluate the effective maximal compression ratio, the dataset was median-filtered along the image stack to produce an image without peaks. 
A background dataset of 1800 such images sparsifies into a 11MB HDF5 file which represents a compression ratio of 136x ! 
Indeed, only 19 pixels were saved per frame and the compressed numerical values are mostly the same, which compresses great with bitshuffle-LZ4.

For ESRF-ID29, where the \textsc{Jungfrau} 4M can operate close to 1 kHz, the pedestal+gain pre-processing convert 16-bits integers into 32-bits floating point values, doubling the bandwidth for data saving.
The detector outputs the data via $8\times$ 10Gbit/s network links and the storage is performed via a single 25 Gbit/s link, making a minimum compression ratio of 6.4x.

In production conditions, ID29 users have to trade between:
\begin{itemize}
    \item Detector speed: operate at lower speed (231 Hz) to be able to save the pre-processed data.
    \item Energy resolution: floating data compress badly, so they are better stored as integer, where the number of ADU per photon can be tuned. With 1 ADU per photon, one gets the best compression rates but one looses the sub-photon energy resolution of the \textsc{Jungfrau} detector.
    \item  Use the burst mode: acquire only shorter datasets and make pauses to let the data reach the disk in the central storage.
    \item Discard pixel of lower intensity: this is a new possibility offered by this algorithm. 
    A cutoff at $1.5\sigma$ should already providing the 6.4x compression ratio needed (Fig. \ref{distribution}).
\end{itemize}

\subsubsection{Compression speed:}

The compression speed has been measured on the computer designed for online data-reduction of the \textsc{Jungfrau} detector \cite{sri2021}: 
an IBM AC922 using two Power9 processors and two Nvidia Tesla V100 GPU. 
The sequential execution of the code on the GPU takes about 4 ms to process one image and uses one single CPU-core and a GPU. 
In production conditions, two such computers are diving each two GPUs, allowing to use the detector  at its nominal speed close to 1kHz.
The main bottleneck remains the networked saving of the different files: preprocessed files, sparse files, peak positions, accumulated frames, ... not all can be saved at all time when operating at full speed.

\subsubsection{Decompression speed:} 
The decompression of those data should typically be performed on a standard workstation (here running two Intel Xeon Gold 6134 CPU @ 3.20GHz): the reconstruction speed is found to take 30 s for the full dataset, while writing of the densified dataset (with bitshuffle-LZ4 compression) takes 45s. 
Densification is thus faster than writing on disk.
The reading time of the input sparse dataset is negligible (<2s).

\subsubsection{Quality of the restored dataset:} 
The densified dataset was processed via XDS and the summary indicator for the quality of the results are compared with the one coming from the reduction of the original dataset. 
Since those integrator are measured on integral peaks with $I/\sigma>3$ and the sparsification was performed
with a cut-off of 2, those result should be almost unaffected, which is confirmed in the table \ref{xds_summary}.

\begin{table}
\label{xds_summary}
\caption{Quality indicators after peak integration and averaging using XDS \cite{xds}. 
Lysozyme (HEWL) dataset provided by Dectris for advertising their Eiger-4M detector \cite{lysozyme}.}.
\begin{tabular}{|c|c c c|c c c|} 
\hline
Indicator & \multicolumn{3}{c|}{Initial dataset} & \multicolumn{3}{c|}{Lossy compressed dataset ($2\sigma$)} \\ 
          & 2.91\AA & 2.06\AA & all & 2.91\AA & 2.06\AA & all \\
\hline
Completeness                                        & 98.8& 90.8 & 93.8\% & 99.8& 90.6 & 93.5\% \\ 
$R_{obs}=\frac{\sum |I_{h,i}-I_{i}|}{\sum I_{h,i}}$ & 9.9 & 57.3& 12.5\% & 9.2 & 61.2&  11.4\%\\ 
$R_{expected}$                                      & 8.8 & 73.2& 15.0\% & 8.2 & 68.7 &  12.1\%\\
$R_{meas}$ \cite{Rmeas}  &10.3 &61.2& 13.2\% & 9.6 & 65.5 & 12.0\%\\
$CC_{1/2}$ \cite{cc1/2}  & 99.7 &94.0 & 99.7   & 99.6 & 95.4 & 99.7  \\
$<I/\sigma>$               & 25.80 & 5.39 & 10.52  & 26.33& 4.09 & 10.17 \\
\hline
\end{tabular}
\end{table}
Of course those data were collected on a test sample with very intense signal, but it demonstrates the algorithm does not destroy the signal. 
But with more challenging samples, exhibiting lower $I/\sigma$, the threshold for picking pixels has to be lower to ensure all pixels relevant for subsequent analysis are actually preserved. 
Unless, and as described in \citeasnoun{veto_2024}, this sparsification would be detrimental for the quality of the reduced data.

\subsection{Influence of the sparsification on the quality of serial-crystallography data acquired with Eiger detector}
\label{nicolas} 

Tiny crystals of egg white lysozyme (HEWL) + Gadolinium were deposited on a SiN membrane and this membrane was scanned on the ID30A3 beamline at the ESRF (\textsc{massif}3) using a Eiger 4M detector.
The dataset consists of 11637 frames of  images, out of which 11512 were properly indexed. 

\subsubsection{Quality indicator degradation:}
\begin{figure}
\label{Rfree}
\begin{center}
\includegraphics[width=9cm]{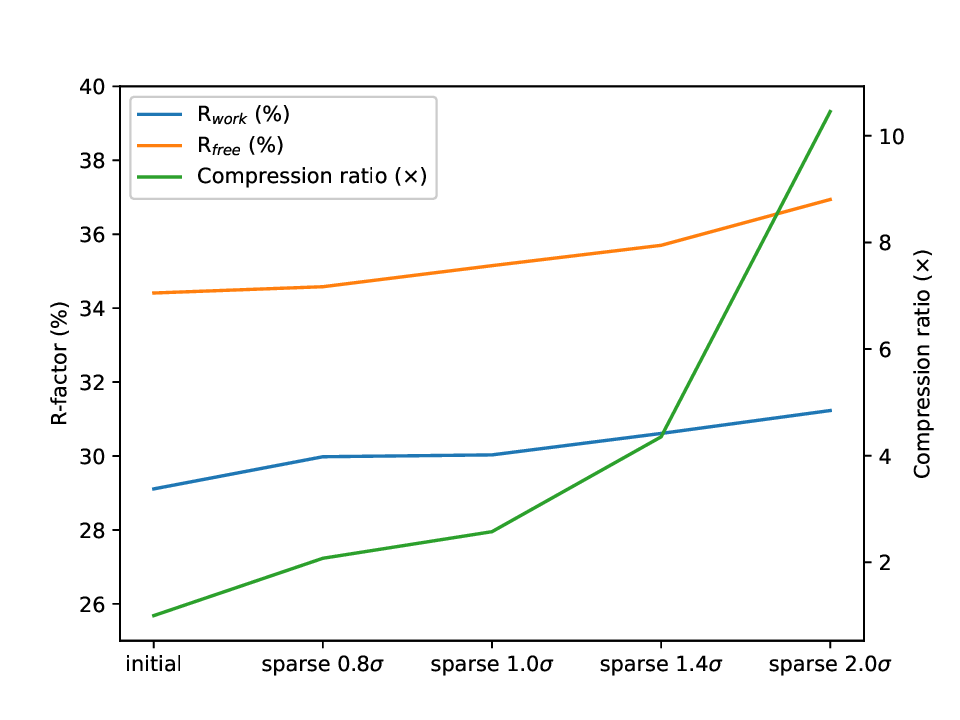}
\caption{Degradation of R$_{work}$ and R$_{free}$ crystallographic quality indicators on the integral dataset of HEWL+Ga (11k frames) and actual compression rates, when increasing the levels of sparsification.}
\end{center}
\end{figure}

The R$_{free}$/R$_{work}$ quality indicators \cite{Rfree} were calculated for the initial dataset and compared with the same dataset sparsified at 0.8, 1.0, 1.4 and $2.0\sigma$ and subsequently re-densified (figure \ref{Rfree}). 

As expected, the crystallographic quality indicator show a gradual degradation with increasing sparsification but R$_{free}$ shows little degradation with little sparsification ($0.8\sigma$, compression ratio of $2\times$).

\subsubsection{Quality indicators as function of the resolution shell:}
\begin{figure}
\label{nicolas_R}
\begin{center}
\includegraphics[width=9cm]{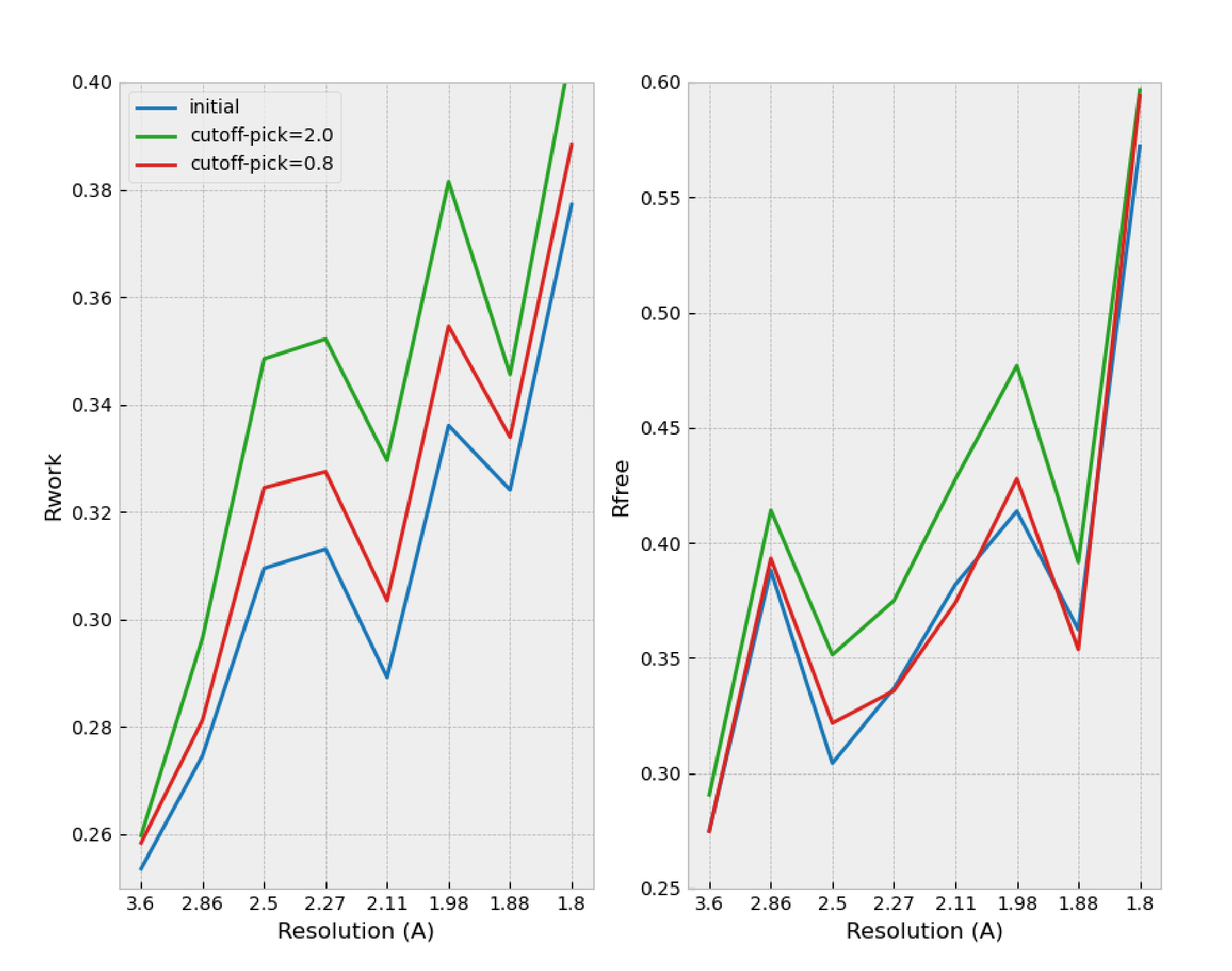}
\caption{R$_{work}$ and R$_{free}$ crystallographic quality indicators at different resolution shells obtained from the complete dataset (11k frames) of HEWL+Ga. The quality of the sparsified data (at $0.8\sigma$ and $2.0\sigma$) are compared with the initial dataset.}
\end{center}
\end{figure}

The same R$_{work}$ and R$_{free}$ quality indicators are reported in figure \ref{nicolas_R} as function of the resolution shell in order to monitor if the degradation is uniform among shells or if it affects mostly the outer shells.
Those indicators compare the inital dataset with the sparsified ones at $0.8\sigma$ and $2.0\sigma$ which were subsequently re-densified for the analysis.

Both R$_{work}$ and R$_{free}$ exhibit degradation as expected and this degradation is rather uniform over all shells, it does not affect more the outer-shells.
The degradation of R$_{free}$ at moderate sparsification ($0.8 \sigma$) is once again very limited.

\subsubsection{Ability to phase sparse data and quality of the density map:}
\begin{figure}
\label{nicolas_A}
\begin{center}
\includegraphics[width=9cm]{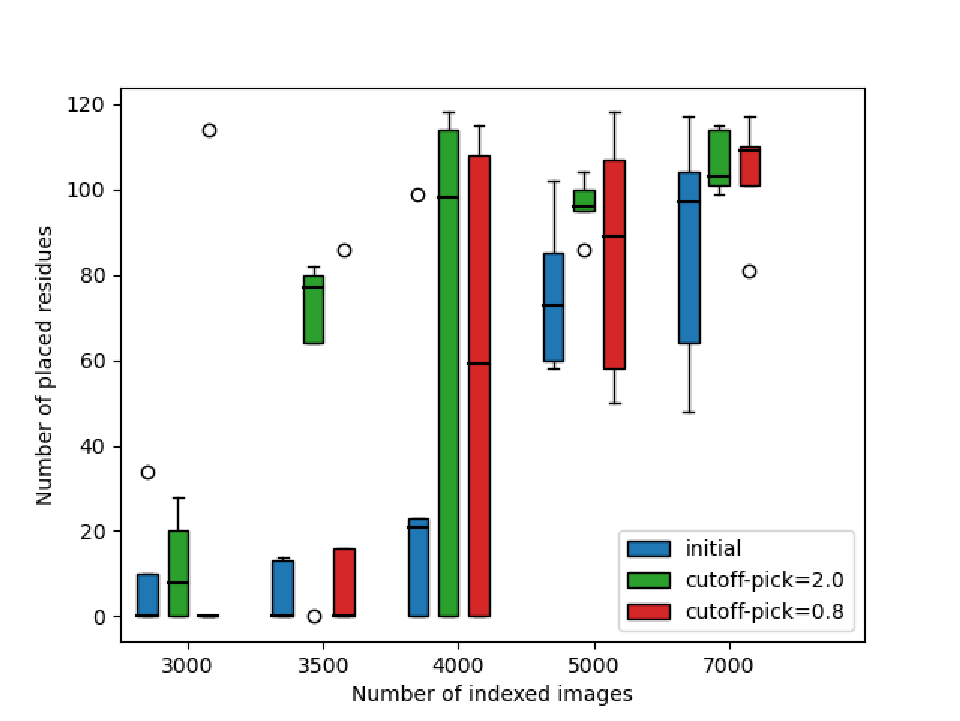}
\caption{Influence of the sparsification (at $0.8\sigma$ and $2.0\sigma$ vs initial dataset) on the ability to phase a the HEWL+Ga protein with an artificially reduced number of frames, in order to limit the strength of the anomalous signal.}
\end{center}
\end{figure}

One of the main concern with sparsification is that it may degrade the weak anomalous signal which is precious for phasing.
The HEWL+Ga dataset was truncated at different length (from 3000 to 7000 out of the 11512 frames) in order to "artificially" decrease the anomalous signal strength in the dataset.
The figure \ref{nicolas_A} shows the number of residues which were properly placed in the electron density map with an automatic procedure.

The sparsified dataset does not show less residues properly placed after the procedure, it even looks marginally better than the initial dataset, especially for larger compression factor where the dataset could be phased with as little as 3500 frames. 
With Eiger detector data, a sparsification at $0.8\sigma$ offers a nice $2\times$ extra compression without noticeable degradation of the quality of the processed data.

\subsection{Influence of the sparsification on the quality of serial-crystallography data acquired with \textsc{Jungfrau} detector}
\subsubsection{Dataset description:} 

Small crystals of NQO1 sample \cite{NQO1}, complexed with NADH, were collected at ESRF-ID29 using a fixed-target sample environment and the \textsc{Jungfrau} 4M detector (PDB: 8RFM). 
The complete dataset represents 574k frames (5 TB) out of which 25809 frames were selected with nice peaks (4.5\% of the total).
The indexation rate for dense data, and two sparse dataset (cut-off at 1.0$\sigma$ and 1.4$\sigma$) are reported in table \ref{t-NQO1} and the crystallographic quality indicators are reported in the figure \ref{f-NQO1} in reciprocal space and figure \ref{r-NQO1} in real space. 

\begin{table}
\label{t-NQO1}
\begin{center}
\caption{Indexation statistics of the NQO1 dataset (25809 frames, 4.5\% of the complete dataset)}.
\begin{tabular}{|c|c | c | c |} 
\hline
       & dense & sparse $1.0\sigma$ & sparse $1.4\sigma$ \\ 
\hline
Disk space & 166 MB & 52 MB & 32 MB \\
Compression & 1$\times$ & 3.2$\times$ & 5.2 $\times$ \\
Frames indexed  & 22874 & 20176 & 20384 \\
Indexation rate & 88.6\%& 78.2\% & 79.0\%\\
\hline
\end{tabular}
\end{center}
\end{table}

\subsubsection{Data statistics:}

The processing was performed with \textit{CrystFEL} \cite{CrystFEL} (version 0.11) with the geometry of the detector optimized with the \textit{millipede} procedure \cite{millepede, millepede2}. 
Indexing was performed with the default parameters of \textit{xgandalf} \cite{xgandalf} based on peak-position by \textit{peakfinder8} \cite{Cheetah2014}.

\begin{figure}
\label{f-NQO1}
\includegraphics[width=12cm]{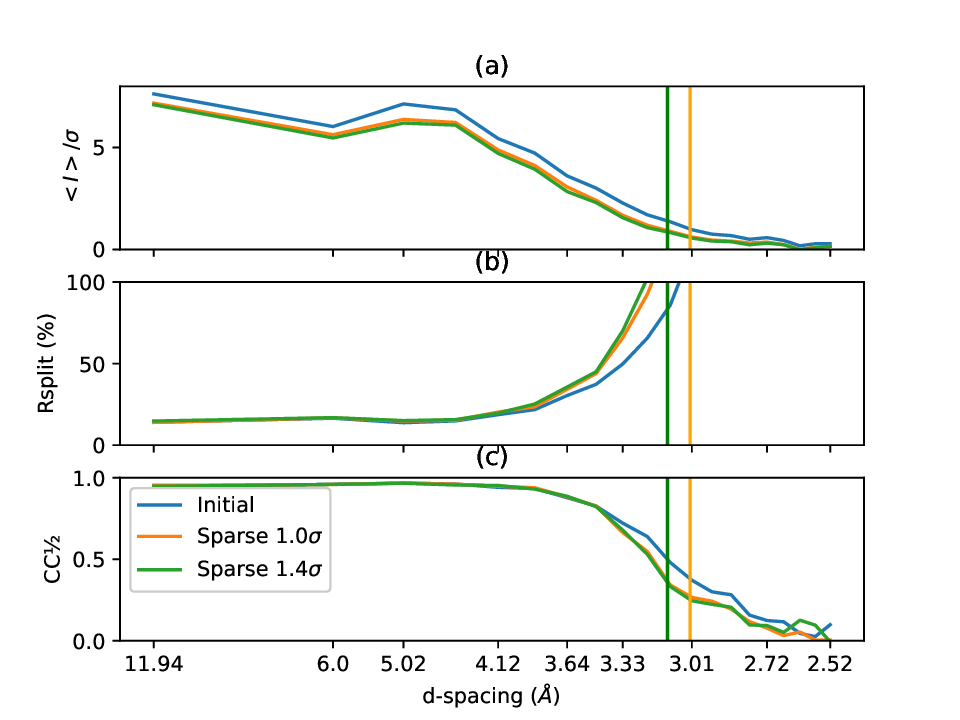}
\caption{Comparison of crystallographic quality indicator as function of the resolution shell for sparsified data ($1.0\sigma$ in orange and $1.4\sigma$ in green) in comparison to the initial dense data (in blue):
(a) signal/noise ratio, (b) R$_{split}$ and (c) CC$_{1/2}$.
The two vertical lines indicate the resolution shell at which the signal/noise ratio drops below the sparsification threshold, i.e. the limit at which all signal is expected to be lost.}

\end{figure}

From the $<I>/\sigma$ curve, figure \ref{f-NQO1}a, one can see there is a systematically lower SNR from sparsified data in comparison with the initial dataset, regardless to the resolution shell. 
This could be due to a too large amount of noise added when densifying the data, especially that this dataset was acquired (and processed) at the full energy resolution of the \textsc{Jungfrau} detector, here 485 ADU/photon (in comparison, an Eiger has 1 ADU/photon).
The SNR curve obtained on uncompressed data was used to assess the resolution shell at which there is no more signal expected to be saved, those are the two vertical lines in green ($1.4\sigma$) and 
orange ($1.0\sigma$). 
On the right of those lines, there is supposed to be no more signal, while on the left, there is supposed to be no degradation of the signal in an ideal case.
The evolution of R$_{split}$ and CC$_{1/2}$, plotted in figure \ref{f-NQO1} b and c, show a degradation 
of the quality which is much earlier, by half an Angstrom, in comparison with what would have been expected. 
\subsubsection{Refinement statistics:}

\begin{figure}
\label{r-NQO1}
\includegraphics[width=12cm]{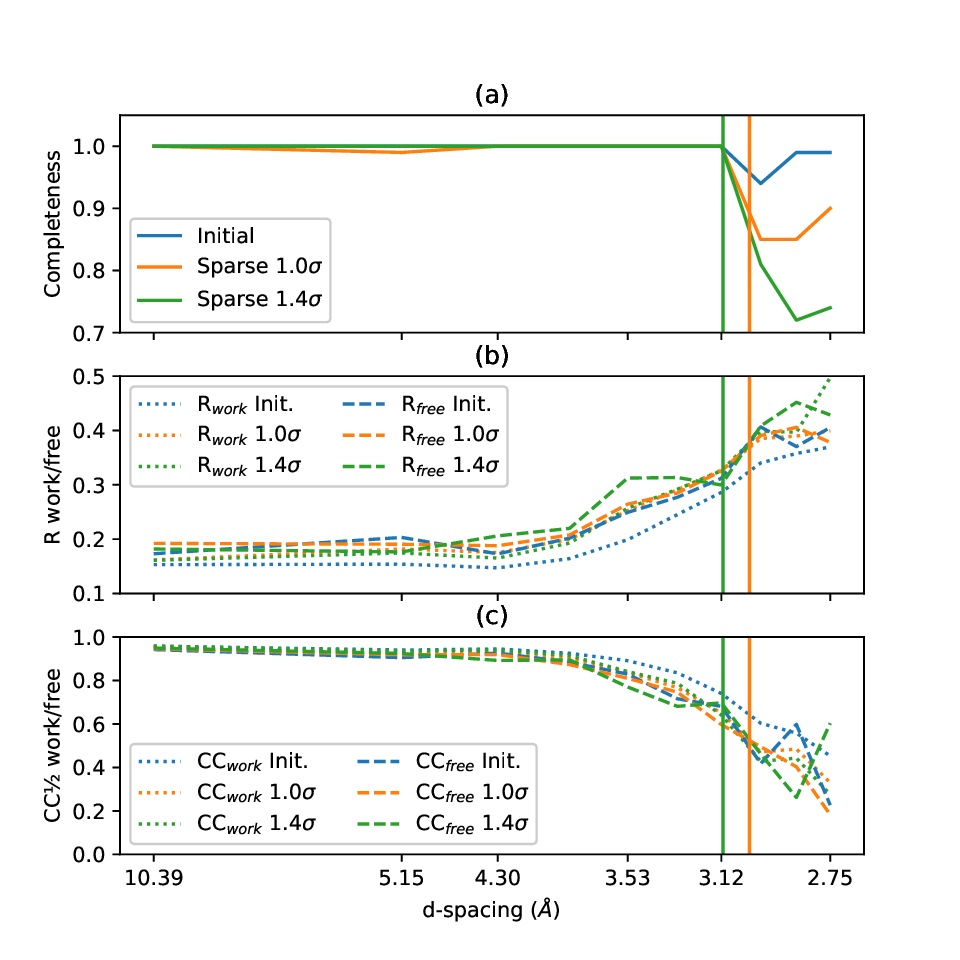}
\caption{Comparison of crystallographic quality indicator in real space after refinement using \textit{phenix.refine} as function of the resolution shell for sparsified data ($1.0\sigma$ in orange and $1.4\sigma$ in green) in comparison to the initial dense data (in blue):
(a) Completeness of the dataset, (b) R$_{work}$ (dotted) and R$_{free}$ (dashed) and (c) CC$_{1/2}^{work}$ (dotted) and CC$_{1/2}^{free}$.
The two vertical lines indicate the resolution shell at which the signal/noise ratio drops below the sparsification threshold, i.e. the limit at which all signal is expected to be lost.}
\end{figure}

Those data were finally refined in real-space using \textit{phenix.refine} \cite{phenix} until a resolution of 2.7\AA, which is noticeably better than what can be expected from the sparse data (limited to 3.0\AA ~and 3.1\AA ~for $1.0\sigma$ and $1.4\sigma$, respectively). 
The results are summarized in figure \ref{r-NQO1}.

The figure \ref{r-NQO1}a contains the completeness of the dataset as function of the resolution shell. 
It confirms the sparsification algorithm works as expected, and that the signal starts to degrade where the SNR drops bellow the picking threshold. 
The figure \ref{r-NQO1}b superimposes the R$_{work}$ in dotted lines (full dataset) in comparison with the R$_{free}$ where the fit is performed on 95\% of the dataset and the quality assessment is performed on the remaining 5\% (dashed line).
Thus R$_{work}$ is always expected to be better than R$_{free}$.
While the R$_{work}$ for the initial dataset looks much better than the sparsified version, the degradation is very limited for the R$_{free}$, especially between the initial and the sparsified dataset at $1\sigma$.
The same is observed for the CC$_{1/2}$ in figure \ref{r-NQO1}c: the initial dataset shows a clear degradation between the \textit{work} and the \textit{free} version, but this degradation is less important for the sparsified version.
The CC$_{1/2}^{free}$ indicators hardly degrades between the initial dataset and the two sparsified ones, confirming the ability to solve the protein structure from sparse data.

\subsection{Conclusion on sparsification}
The sparsification algorithm presented here is generally applicable to any kind of single crystal diffraction experiment where the background is isotropic. 
This excludes notably diffuse scattering experiment but it is generally applicable to small molecules crystallography and macro-molecular crystallography and even serial crystallography, when the images are nicely centro-symmetric and shadows are properly masked out.
The additional compression offered by sparsification is especially interesting for serial crystallography where datasets of millions of frames are collected per experiment.
As any lossy compression, the user has still the responsibility of choosing wisely the threshold level as it will limit later-on the quality of the results one may extract from those data. 
This is of crucial importance for macromolecular crystallography where valuable information is still present in reflection with SNR less than one \cite{cc1/2}.
We have demonstrated that a sparsification at $0.8\sigma$, which has an actual compression of $2.0 \times$ on Eiger data ($2.6 \times$ for Jungfrau data), preserves nicely the electron density map and is hardly distinguishable from the uncompressed data.
The sparsification at $1.0\sigma$, offering a $2.6\times$ compression on Eiger data ($3.2\times$ for Jungfrau), preserves enough signal to refine a protein in with very limited degradation of the R$_{free}$ and CC$_{1/2}^{free}$ quality indicators.
The Jungfrau data show systematically larger compression rate for sparsification because data was collected with 485 ADU/photon (and is tunable) while Eiger detector always operate at 1 ADU/photon (i.e. it has less noise).

In comparison with ROIBIN-SZ \cite{roibin}, which preserves all pixels in the neighborhood of identified peaks and stores the background heavily binned, the sparsification presented here does not require a complete peak-search thus, it is simpler, with fewer parameters to be tuned, and is able to save any pixel which is intense enough.
Nevertheless, peak-finding is of crucial important for identifying Bragg-peaks, one of the first stage of any data reduction pipeline.

\section{Application to peak-finding for serial crystallography}
A classical way of pre-processing serial-crystallography data is to shrink the amount of data by sieving-out empty or bad frames, only keeping the frames which deserve processing. 
This is the role of the "veto-algorithm".

The sigma clipping algorithm provides us with the background (average and deviation) and is used to pick pixels which are likely to be part of Bragg peaks, like peakfinder8 \cite{Cheetah2014} does. 
For this, several additional checks are performed on a local neighbourhood which is a small square patch (typically 3x3 or 5x5 pixels, user defined):
\begin{itemize}
\item Is the considered pixel the maximum of the local neighborhood ?
\item Are enough pixels of the local neighborhood satisfying the SNR condition? (user defined parameter)
\end{itemize}

\begin{figure}
\label{silx}
\includegraphics[width=12cm]{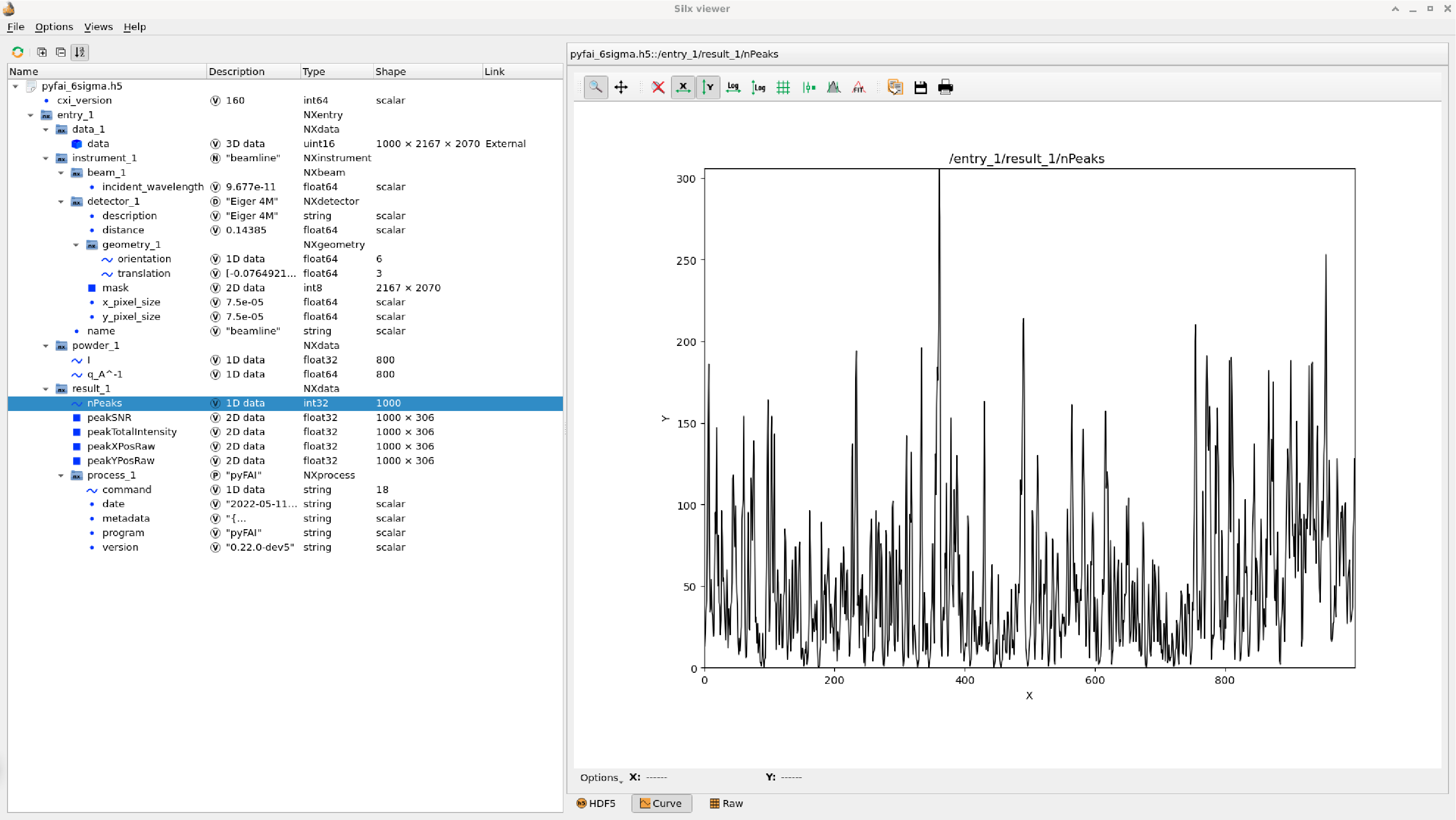}
\caption{Peak-picking CXI-file produced by pyFAI and visualized with the \textit{silx viewer} \cite{silx}.
The left-hand side contains the HDF5 tree structure while the right-hand side presents the default plot with the number of peaks found per frame.}
\end{figure}

For each peak, the coordinates of the centroïd, the sum of data and its propagated deviation are recorded and reported. 
Those peak-position are saved into a HDF5 file (represented figure \ref{silx}) following the CXI format \cite{cxi} which can be read from CrystFEL \cite{CrystFEL}.
CrystFEL allows to swap peak-picking algorithms \cite{zaefferer, Cheetah2014, robustpeakfinder} and indexing tools \cite{xds, mosflm, taketwo, xgandalf, pinkindexer}.

The serial crystallography beamline at the ESRF (ID29) uses a LImA2 monitor \cite{lima2} as visualization tool.
It is inspired from  NanoPeakCell \cite{nanopeakcell} for online visualization and feeds back information to check if peaks found correspond actually to the crystal lattice expected for the sample.
We will first compare the peak-picking algorithm with some reference implementation on a single frame before evaluating the quality of the picked points on a serial-crystallography dataset.

\subsection{Comparison of picked peaks}
\begin{figure}
\label{peakfinder}
\includegraphics[width=12cm]{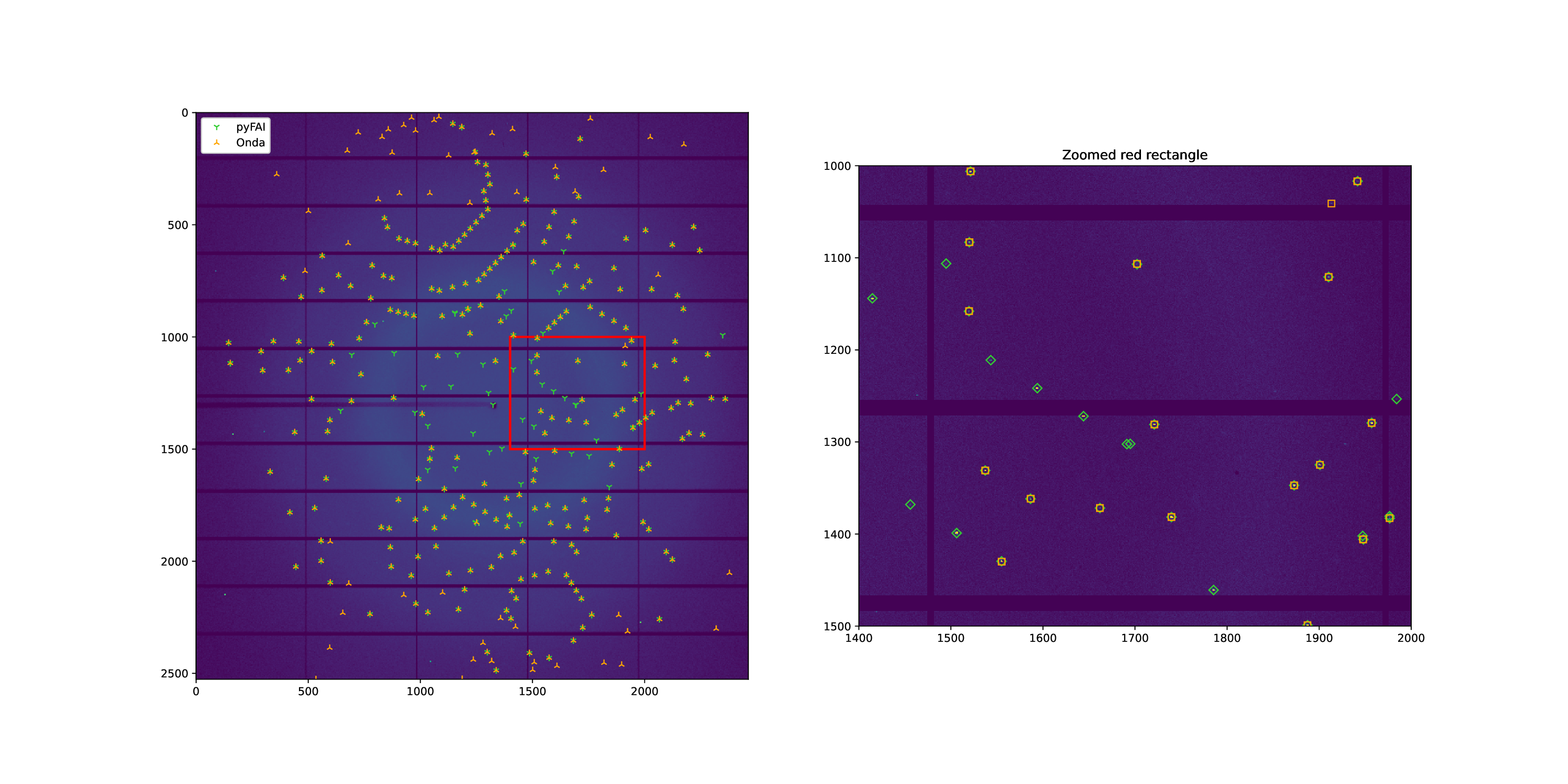}
\caption{Comparison of the reference \textit{peakfinder8} interfaced with Onda (in orange, execution time 300ms) and the version from pyFAI (in green, execution time 10ms) on top of a Pilatus 6M diffraction frame of an insulin crystal.
The subplot on the right is a close-up to the red rectangle.}
\end{figure}

Figure \ref{peakfinder} presents the comparison between the original \textit{peakfinder8} described in \citeasnoun{Cheetah2014}, interfaced in Python via OnDA \cite{onda} and the version implemented into pyFAI on the same Pilatus 6M image, already used in figure \ref{fig2}. 

\begin{figure}
\label{peak_per_ring}
\includegraphics[width=12cm]{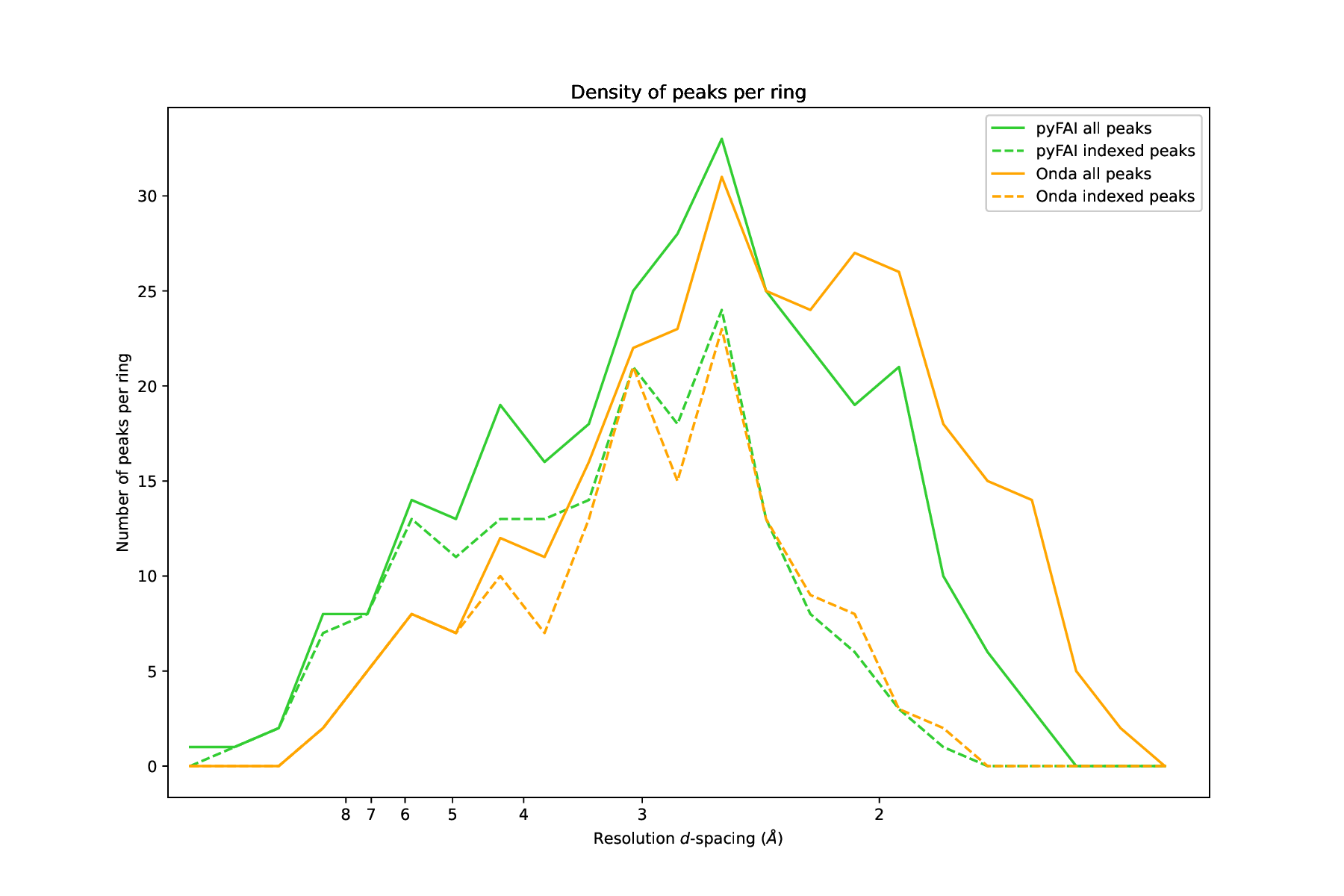}
\caption{Number of peaks found in the different resolution shells for the \textit{peakfinder} implemented in pyFAI and in Onda. 
The width of each radial-shell is two inverse nanometer in $q-$space. 
The dashed lines represents the number of those peaks which have successfully been indexed, i.e. located at less than two pixels away from an expected reflection (calculated by \textit{xgandalf}). }
\end{figure}

In figure \ref{peakfinder}, most peaks found by both implementation match and correspond to Bragg reflections, the close-up on the right allows to visualize Bragg spots in the image.
There are more green peaks (found by pyFAI) closer to the beam center while more orange peaks (found by Onda) are located in the outer-shell.
This plot was made with a minimum SNR of 3 and a noise level of 1 since the Pilatus detector is Poissonian.
Peaks were registered if four pixels meet the SNR criterion in a 5x5 pixels patch around the peak.
Those parameters have been tuned to obtain a comparable number of peaks with both implementations: 290 with pyFAI and 293 with Onda.
The similarity of those figures allows a direct comparison of peaks found per resolution shell, histograms which are plotted in figure \ref{peak_per_ring}.

The figure \ref{peak_per_ring} presents the histogram of $q$-values (modulus of the scattering vector) of the peaks found with different methods, i.e. the number of peaks having a given q-value.
For readability, those histograms (bin width: $2 nm^{-1}$) have been represented as plots with the horizontal axis labeled in d-spacing ($q=2\pi/d$).  
The analysis of those histograms confirms that the implementation in pyFAI is getting more points closer to the beam-center while the reference implementation is picking more point at larger q-values.
This is likely due to the curvature of the Debye-Scherrer ring: the original version of peakfinder8 evaluates the variance in a neighborhood defined by some radius around the point of interest and the variance is higher close to the beam-center due to the important curvature of those rings.
On the opposite, pyFAI knows about this curvature and measures the variance along the ring.
Points picked by Onda at larger q-values do not look like Bragg-peaks but this could be a side-effect of the parameter tuning to get the same number of peaks for both algorithms.

The same figure presents with dashed lines the number of peaks which are coincident (within 2 pixels) with expected reflections (after indexing using \textit{xgandalf}).
It demonstrates that the additional peaks found by pyFAI in the inner-shell ($d>3$\AA{}) are consistent with Bragg-peaks, thus valuable for indexing, while the additional peaks found by Onda in the outer shell ($d<2$\AA) are not Bragg-peaks, and thus are detrimental.

A word on performances: the Python binding in Onda to the peakfinder8 algorithm from Cheetah runs in 180ms on a high-end server CPU (AMD Epyc 7262) and 300ms on a workstation (Intel Xeon E5-1650 v4). 
The version available in pyFAI was designed in OpenCL \cite{opencl_khronos, opencl, pyopencl} and runs best on GPUs: 30ms on a AMD Vega 56 and in 10 ms on a Nvidia RTX A5000. 

\subsection{Quality of the peakfinder on serial crystallography data}
\label{quality-pf}
The quality of peaks extracted with this algorithm is evaluated on a serial-crystal\-lography dataset. 
A subset of 1000 frames of the dataset used in section \ref{nicolas}
was used as a probe and was indexed with the $indexamajig$ tool from CrystFEL. 
Since all frames show Bragg-peaks (figure \ref{silx}) the number of indexed frames can be seen as a quality indicator of the peak-picking algorithm used, when all other parameter remain unchanged.
The indexation was performed with the \textit{xgandalf} algorithm \cite{xgandalf} with default settings from CrystFEL v0.10.1 and was provided with the cell parameter: tetragonal $a=b=78.77\text{\AA}; c=39.04\text{\AA}; \alpha=\beta=\gamma = 90^o$.
The table \ref{crystfel} compares the number of frames properly indexed with the different picking algorithms available in CrystFEL and with the algorithm presented here. 

\begin{table}
\label{crystfel}
\caption{Indexation rate obtained with \textsc{xgandalf} \cite{xgandalf} from peak positions extracted with different picking algorithms available from CrystFEL \cite{CrystFEL} on a subset of 1000 frames of microcrystals of Lysozyme (HEWL+Ga) collected with an Eiger 4M at ESRF-ID30A3.}
\begin{center}
\begin{tabular}{|c|c|c|c|c|} 
\hline
Peak-picking  & \multicolumn{2}{c|}{\textsc{xgandalf} (default)} & \multicolumn{2}{c|}{\textsc{xgandalf} (fast)}\\
method & Index. rate & run-time & Index. rate & run-time \\ 
\hline
Zaef \cite{zaefferer} & 10.0\% & 2178 s & 10.0\%& 430 s\\
PeakFinder8 \cite{Cheetah2014} & 49.5\%& 10397 s &48.5\% & 1757 s\\
PeakFinder9 \cite{FDIP} & 44.2\%& 8328 s&43.5\%&1436 s\\
RobustPF \cite{robustpeakfinder} & 22.4\%& 6314 s& 21.2\%& 1628 s \\
pyFAI (this contribution) & 49.7\%& 9325 s& 49.2\% &1595 s\\
\hline
\end{tabular}
\end{center}
\end{table}

Since the Eiger detector is a counting detector, the global threshold for algorithms \textit{zaef} and \textit{peakfinder8} had to be lowered (to 50, which is the maximum of the background signal on any frame) and the the default SNR value was used for \textit{zaef}, \textit{peakfinder8}. 
The same SNR value of five was used for \textit{pyFAI}.
Default parameters were used for \textit{PeakFinder9} and \textit{RobustPeakFinder} algorithms.
The reported run-time correspond to the execution time on a single core of an Intel Xeon Gold 6134.

The indexing rate obtained with the algorithm from pyFAI is in par with the reference implementations like \textit{peakfinder8} or \textit{peakfinder9} available from CrystFEL. 
Since time is mostly spent in indexing, sets of peaks which are simpler to index present lower run-time as fewer retry were needed.
Retries can be deactivated but they increase significantly the success of indexing. 
Table \ref{crystfel} presents also the indexing performances when using the option \textit{--xgandalf-fast-execution} from \textit{indexamajig}, which is five to six times faster and exhibits a limited degradation of the indexing rate.
The peak extraction in pyFAI is about as fast as the sparsification, so it can be used online to perform the pre-analysis and provide peaks to NanoPeakCell. 
Nevertheless the executable `peakfinder` available from pyFAI (offline tool) has a total execution time which is much larger: about 30 s for 1000 frames, most of which is spent in reading and writing the different HDF5 files.

\subsection{Peak count as $veto$ algorithm}
Since the background extraction and peak finding are performed in real-time on the serial-crystallography beamline ID29 at ESRF, the information about the number of Bragg-spots can be used to assess the quality of each individual image and the acquisition system can decide to discard the frame depending on the number of peaks and a live-adjustable threshold. 

\begin{figure}
\label{veto-fig}
\begin{center}
\includegraphics[width=9cm]{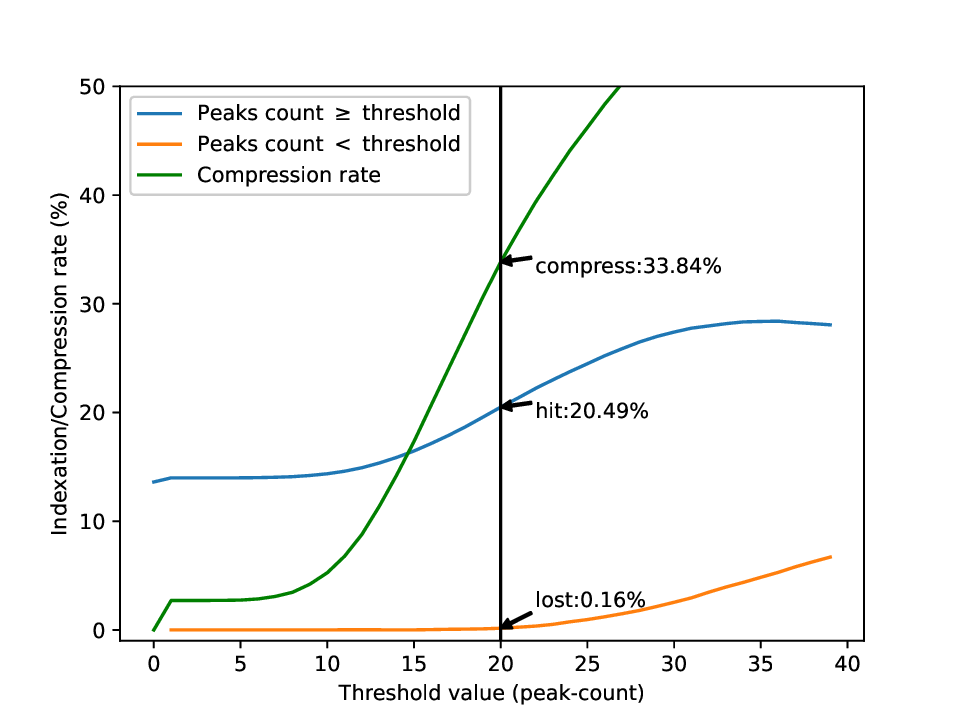}
\caption{Indexation rate of frames which would be considered as "hit" or "non-hit" as function of the peak-count threshold. 
The green curve presents the disk space which could have been saved.
The sample dataset consists of 80000 frames of lysozyme micro-crystals indexed off-line with Xgandalf in CrystFEL.}
\end{center}
\end{figure}

Since the beginning of operation of ESRF-ID29, in 2022, the beamline operated with the $veto$ algorithm deactivated, the number of peaks found was just recorded for future exploration. 
The figure \ref{veto-fig} presents the indexation rate of "hit" (in blue) and "non-hit" frames (in orange) when changing the threshold for the minimum peaks-count per frame. 
The expected compression rate is displayed in green.
The dataset consists of 80000 lysozyme micro-crystals between two mylar films, raster scanned with a x-ray beam of 11.56 keV at ESRF-ID29.
The offline analysis has been performed with CrystFEL (v0.11.0) using `peakfinder8` and `xgandalf` as indexer. 

Since the \textsc{Jungfrau} is an integrating detector, the detector has relatively more background noise than a photon counting detector: here, data were saved with 8 ADU/photon, making the compression rate of hit and non-hit frames similar.
In this example, discarding frames with less than 20 peaks would have allowed to save one third of the network bandwidth and disk space, loosing only 0.16\% of frames index-able. 
The indexation rate of actually recorded frames also increased to 20\%.
It is noticeable the hit-rate was especially high in this experiment with 84\% of frames showing diffraction peaks.
One would not expect a user-experiment to have such high concentration of crystals and in normal conditions, the expected compression ratio should be higher.
The $veto$ algorithm having proved its robustness, it is now activated for most experiments, thanks to graphical helper tools to asses the optimal threshold during the experiment and allowing to set those parameters on the fly for real-time processing.

\subsection{Limitations}
There is a strong sensitivity of the indexation rate with data from the \textsc{Jungfrau} detector, related to the description of the detector in CrystFEL and in pyFAI.
The \textsc{Jungfrau} 4M detector is built from 8 modules manually assembled and exhibits some residual misalignment, on the order of a few pixels and miss-behaving pixels.
This misalignment is much larger than what is commonly encountered with Eiger detectors from Dectris, where misalignment is usually less than one pixel in size (75 µm).
On Eiger detector data, where the detector is defined as a single rigid module, the indexation rate is fairly independent of the peak-picking algorithm as described in \ref{quality-pf}.
This means the peak positions provided by the veto-algorithm are suitable for indexing and this saves the reading of the complete frame if it cannot be indexed.
With data from the \textsc{Jungfrau} detector, the indexing rate was lower than with certain peak-finders algorithms integrated into CrystFEL. 
Work is ongoing to convert the geometry description from pyFAI, where every pixel is independent, to the geometry used in CrystFEL, and vice-versa. 
We found this difference of indexation was related to the description of the mask and of pixel position in different software.
Until this issue is addressed, the safest is to re-extract peak positions for indexing using one of CrystFEL's provided peak-finding algorithms. 

\section{Conclusion}

Background analysis of single crystal diffraction images can be implemented efficiently using iterative azimuthal integration and allows the separation of the signal originating from Bragg-peaks from isotropic background.

A lossy compression algorithm for diffraction frames, called sparsification, has been built on top of this signal separation, with the average background saved on the one side and the position and the intensity of most intense pixels (probably belonging to peaks) on the the other.
The quality of the compression has been demonstrated on macro-molecular rotational and serial crystallography data. 
The degradation of the signal has been monitored after a compression-decompression cycle, both in reciprocal space using crystallographic quality indicators and after Fourier transform in direct space where the number of residues placed automatically was checked.
The degradation of the signal was also monitored as function of the threshold and sparsification at $0.8\sigma-1.0\sigma$ still enables to reconstruct the molecular structure of proteins, both with Eiger and \textsc{Jungfrau} detectors.

The second application is a peak-finder for serial crystallography which locates peak positions in real time and can be used as veto algorithm to discard images without (enough) diffraction peaks. 
The peaks picked have been evaluated against state of the art peak-finder algorithms like \textit{peakfiner8} and the results were comparable in quality while much faster thanks to the usage of GPU.  
This \textit{veto-}algorithm is now used in production at the ESRF serial crystallography beamline (ID29) to save storage space.

One of the strength of this peak-finer algorithm is that it is optimized to work on GPU and thus ideally suited to be coupled with the next generation of crystal indexer \cite{toro} which is running on the same kind of hardware, allowing to save two memory-transfer per frame and maybe achieve real-time integration of serial crystallography data.



\ack{Acknowledgements:}
The authors would like to thank Andy G\"otz for the management of this project and Vincent Favre-Nicolin for his support.
Thanks also to Gavin Vaughan, scientist at Materials beamline at the ESRF,  for the constructive discussion about sigma-clipping versus median filtering.
We would like to thank also Jonathan P. Wright and Carlotta Giacobbe for offering us the ability to test those algorithms on small molecule data and validate the concept of sparsification on the ID11 beamline at the ESRF.
Pierre Paleo and Jerome Lesaint are also acknowledged for the fruitful discussions on numerical methods developed in this document.

\bibliographystyle{iucr}
\bibliography{biblio}

\end{document}